\documentclass[lettersize]{IEEEtran}
\usepackage{amsmath,amsfonts}
\usepackage{algorithmic}
\usepackage{algorithm}
\usepackage{array}
\usepackage{textcomp}
\usepackage{url}
\usepackage{verbatim}
\usepackage{graphicx}
\usepackage{cite,color,tcolorbox}
\usepackage{amssymb,setspace}
\usepackage{amsthm}
\usepackage{subcaption} 
\usepackage{stfloats}

\usepackage{hyperref}
\usepackage{epstopdf}
\usepackage{geometry}
\usepackage{bm}
\begin{document}
\bstctlcite{IEEEexample:BSTcontrol}

\title{Beyond Fresnel Approximation: Exact Airy Beamforming for Near-Field THz Communications}
\author{Fangjie Shen, Weidong Mei, \textit{Member, IEEE}, Aoran Liu, Peilan Wang, and Zhi Chen, \textit{Senior Member, IEEE}
	\thanks{Fangjie Shen, Weidong Mei, Aoran Liu, Peilan Wang, and Zhi Chen are with the National Key Laboratory of Wireless Communications,
		University of Electronic Science and Technology of China (UESTC), Chengdu 611731, China (email: shenfangjie@std.uestc.edu.cn; wmei@uestc.edu.cn; liuaoran@std.uestc.edu.cn; peilan\_wangle@uestc.edu.cn; chenzhi@uestc.edu.cn).}}  
\maketitle
 
\begin{abstract}  
Line-of-sight (LoS) blockage can severely degrade the performance of conventional focused beams in terahertz (THz) communications. Recently, bending beams (e.g., Airy beams) have attracted increasing attention due to their ability to bypass obstacles by propagating along curved trajectories under near-field propagation conditions. However, existing Airy-beam designs in wireless communications often rely on the Fresnel approximation for the focal distance, which becomes inaccurate at large steering angles or short propagation distances. To address this limitation, this letter proposes a more accurate Airy beamforming design method based on the exact focal-distance expression and the corresponding caustic geometry. Specifically, we jointly optimize the bending curvature, virtual focal distance, and steering angle to maximize the received signal power subject to receiver-alignment and obstacle-avoidance constraints. To tackle the resulting non-convex problem, we first use geometric preprocessing to narrow the feasible range of the bending curvature. Then, a two-layer optimization framework is developed with the outer and inner layers optimizing the bending curvature and the other two parameters, respectively. Numerical results based on the cascaded geometrical and wave channel model demonstrate that the proposed method achieves higher received signal power than conventional Airy-beam designs and offers complementary advantages over conventional focused beams under different blockage ratios.
\end{abstract} 
 

\setlength{\textfloatsep}{3pt plus 1pt minus 1pt}
\setlength{\floatsep}{3pt plus 1pt minus 1pt}
\setlength{\intextsep}{3pt plus 1pt minus 1pt}
\setlength{\abovedisplayskip}{3pt plus 1pt minus 1pt}
\setlength{\belowdisplayskip}{3pt plus 1pt minus 1pt}
\setlength{\abovedisplayshortskip}{2pt plus 1pt}
\setlength{\belowdisplayshortskip}{2pt plus 1pt minus 1pt}
\predisplaypenalty=0

\section{Introduction}
\label{sec:introduction}
Terahertz (THz) communication is regarded as a promising technology for sixth-generation (6G) wireless networks, because its large available bandwidth can support ultra-high data rates\cite{10045774}. However, due to the extremely high carrier frequency, practical THz links are highly vulnerable to physical blockage. Even a small obstacle in the line-of-sight (LoS) path can cause severe received-power loss or even link outage \cite{9940939, 9903389}.
 
Bending beams have recently emerged as a promising solution to mitigate blockage in near-field THz communications \cite{siviloglou2007accelerating, guerboukha2024curving, droulias2024bending, liu2026bending}. Their curved propagation trajectories enable them to bypass obstacles before reaching the receiver (Rx) under the near-field propagation condition. Their self-healing property further facilitates beam reconstruction after partial blockage. Inspired by these properties, recent works have proposed various bending beamforming methods for wireless communications, such as physics-informed deep learning frameworks \cite{Chen2025APA} and successive convex approximation (SCA) algorithms \cite{liu2025near}. Although effective, these methods respectively entail considerable offline training and real-time computational complexity, thereby hindering their practical implementation. Alternatively, Airy beamforming offers a low-dimensional design approach by only controlling three phase parameters, i.e., bending curvature, virtual focal distance, and steering angle \cite{zhang2026breaking,zhao2026efficient,xie2026performance,zhao2026airy}. Among these works, the authors in \cite{zhao2026airy} derived an approximate closed-form solution for these parameters under the Fresnel approximation (i.e., the second-order Taylor expansion of the focal distance). However, this approximation may introduce phase errors and deviations in the beam trajectory, particularly at large steering angles or short propagation distances. These conditions can arise in typical THz communication scenarios, such as indoor data centers, and degrade the performance of obstacle avoidance.

To address this limitation, this letter proposes an Airy beamforming optimization method based on the exact focal-distance expression. Our goal is to jointly optimize the three parameters of the Airy beam at a transmitter (Tx) to maximize the received signal power subject to receiver-alignment and obstacle-avoidance constraints, as depicted in Fig.~\ref{fig:system_model}. However, this method leads to a highly non-convex optimization problem. To tackle this challenge, we first use geometric preprocessing to narrow the feasible range of the bending curvature. Next, a two-layer optimization framework is proposed: the inner layer derives the optimal focal distance and steering angle for a given curvature using a focal-curve intersection method, while the outer layer optimizes the bending curvature via the Piyavskii–Shubert (PS) search. Numerical results based on the cascaded geometrical and wave channel model \cite{zhao2025terahertz} demonstrate the effectiveness of our proposed method compared with the approximate design in \cite{zhao2026airy} and conventional focused beams.

\section{System Model}
\label{sec:System Model}
\begin{figure}[!t]
    \centering
    \includegraphics[width=0.95\columnwidth]{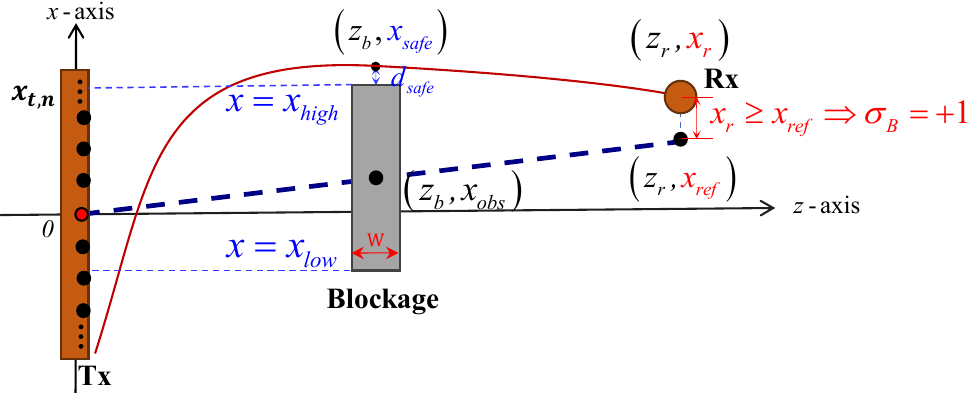}
    \caption{Airy-beam-aided wireless communications.}
    \label{fig:system_model}
\end{figure}
As shown in Fig.~\ref{fig:system_model}, we consider a two-dimensional (2D) near-field THz communication system in the $z$-$x$ plane, where the $z$- and $x$-axes correspond to the propagation and transverse directions, respectively. The Tx employs an $N_t$-element uniform linear array (ULA) aligned with the $x$-axis and centered at the origin. The coordinates of its $n$-th element are $(0,x_{t,n})$, with
\begin{equation}
\label{eq:tx_coordinates}
x_{t,n} = \left( n - \frac{N_t + 1}{2} \right) d, \quad n = 1, 2, \dots, N_t ,
\end{equation}
where $d=\lambda/2$ is the element spacing and $\lambda$ is the carrier wavelength. Let $L=(N_t-1)d$ denote the physical aperture. A single-antenna Rx is located at $(z_r,x_r)$. Furthermore, it is assumed that there is an obstacle in the $z$-$x$ plane between the Tx and the Rx, which is modeled as an ideal absorbing rectangle of thickness $W$. Let $(z_b,x_{obs})$ denote the coordinates of its center, and let $x_{\mathrm{low}}$ and $x_{\mathrm{high}}$ denote its low and high boundaries, respectively. Thus, the obstacle can be expressed as
\begin{equation}
\label{eq:obstacle_mask}
\Omega_{\mathrm{obs}} = \Big\{ (z, x) \ \Big|\ z \in [z_b - W/2, z_b + W/2], \, x \in [x_{\mathrm{low}}, x_{\mathrm{high}}] \Big\} .
\end{equation}
Notably, if $x_{\mathrm{high}} - x_{\mathrm{low}}$ is sufficiently large, the direct Tx-Rx path may be severely blocked. To ensure reliable communications in this case, we consider that the Tx steers an Airy beam to bypass the blockage and communicate with the Rx.

To characterize the field intensity at the Rx, we first model the Tx array as a continuous aperture and denote $x_0\in[-L/2,L/2]$ as any point within it. In an unobstructed free-space scenario, the electric field at any position can be obtained based on the first Rayleigh-Sommerfeld (RS) integral as
\begin{equation}
\label{eq:discrete_RS} 
E(z, x) = \frac{1}{\mathrm{j}\lambda} \int_{-L/2}^{L/2} E(0, x_0) \frac{\mathrm{e}^{\mathrm{j}kr(x_0)}}{r(x_0)} \cos(\theta(x_0)) \mathrm{d}x_0 ,
\end{equation}
where $E(0, x_0)$ represents the initial electric field at the source point $(0, x_0)$, $k = 2\pi/\lambda$ is the wavenumber, $r(x_0) = \sqrt{(x-x_0)^2 + z^2}$ is the exact propagation distance, and $\theta(x_0)$ denotes the angle between the propagation vector and the $z$-axis, with the directional cosine $\cos(\theta(x_0)) = z/r(x_0)$.

\section{Exact Airy Beam Generation and Problem Formulation}
This section first characterizes the exact Airy-beam trajectory and then formulates the parameter optimization problem.

\subsection{Airy-Beam Construction}
\label{sec:discretization}
\label{sec:caustic}
\label{sec:offset_and_limitations}
\begin{figure}[!t]
    \centering
    \includegraphics[width=0.91\columnwidth]{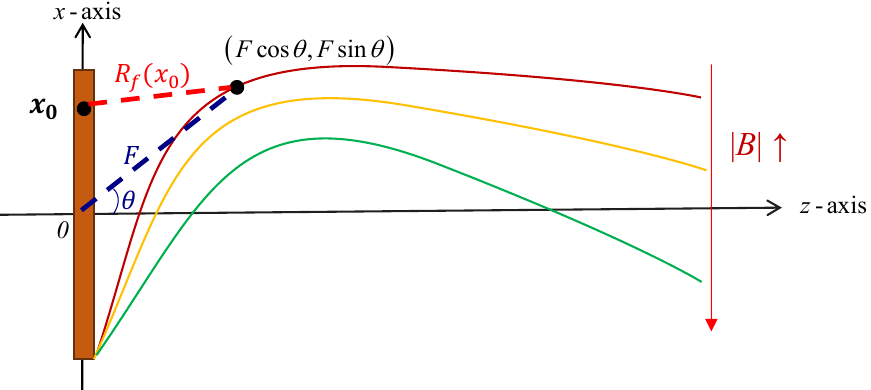}
    \caption{Illustration of the Airy beam parameters.}
    \label{fig:caustic_trajectory}
\end{figure}
Based on the principles of Fourier optics \cite{Efremidis:19}, the phase shifts of the Airy beam can be synthesized by applying a continuous cubic phase mask onto an exact-distance focusing phase, which is given by \cite{phase, Chen2025APA}: 
\begin{equation}
    \label{BF}
    \phi(x_0; B, F, \theta) = \frac{1}{3}(2\pi B)^3 x_0^3 - k R_f(x_0),
\end{equation}
where $B$, $F$, and $\theta$ denote the bending curvature, virtual focal distance, and steering angle of the Airy beam, respectively, as illustrated in Fig.~\ref{fig:caustic_trajectory}. Parameter $B$ controls the bending curvature: a smaller/larger value of $|B|$ produces smoother/sharper turns; $F$ is the virtual focal distance, and $\theta$ is the steering angle. The distance $R_f(x_0) = \sqrt{(F\sin\theta - x_0)^2 + F^2\cos^2\theta}$ measures the exact distance from the source coordinate $(0, x_0)$ to the virtual focal point $(F\cos\theta, F\sin\theta)$. Note that if $B=0$, the Airy beam reduces to a beam focused at the focal point.
 
Based on \eqref{BF}, the trajectory of the bending beam can be analytically derived as the envelope of the rays emitted from the aperture, mathematically defined as a caustic. Specifically, let $\alpha(x_0)$ denote the local launch angle at $x_0$, which can be obtained based on
\begin{equation}
    \label{launch angle}
    \sin \alpha(x_0) = \frac{1}{k} \left. \frac{\partial \phi(x)}{\partial x} \right|_{x=x_0} = \frac{(2\pi B)^3}{k} x_0^2 + \frac{F \sin \theta - x_0}{R_f(x_0)}, 
\end{equation}
and the corresponding ray path is given by
\begin{equation}
    \label{emitted ray}
    x(z, x_0) = x_0 + z \cdot \tan\alpha(x_0).
\end{equation}

By evaluating the geometric envelope of these emitted rays, the exact caustic trajectory can be obtained as \cite{foulaadvand2021flux}
\begin{equation}
    \label{trajectory}
    \begin{cases}
        z_c(x_0) = \dfrac{-\cos^2\alpha(x_0)}{\alpha'(x_0)}, \\[3pt]
        x_c(x_0) = x_0 + z_c(x_0)\tan\alpha(x_0),
    \end{cases}
\end{equation}
for $x_0\in[-L/2,L/2]$, where $\alpha'(x_0) = \frac{\mathrm{d}\alpha(x_0)}{\mathrm{d}x_0}$. 

Note that the caustic in (\ref{trajectory}) defines the outermost boundary of the emitted rays. Since the physical main lobe has a finite beamwidth, its peak intensity is slightly offset from this outermost geometric boundary. By accounting for this discrepancy, the corrected physical peak trajectory is obtained by shifting $x_c(x_0)$ as \cite{zhao2026airy, Efremidis:19}
\begin{equation}
    \label{trajectory peak}
    x_{peak}(x_0) = x_c(x_0) - \left( \frac{-1.0188 \cdot \lambda \cdot z_c(x_0) \cdot B}{\cos^2\alpha(x_0)} \right),
\end{equation}
where the constant 1.0188 originates from the global maximum of the squared Airy function, $\text{Ai}^2(\xi)$, whose primary peak is located at $\xi \approx -1.0188$, and the factor $1/\cos^2\alpha(x_0)$ is introduced to account for the non-paraxial transverse projection.

\subsection{Problem Formulation}
\label{sec:optimization_formulation}
The above bending beam is generated by assuming a continuous phase profile within the transmit aperture. Due to the finite number of transmit antennas, we evaluate the electric field by sampling the aperture at the antenna coordinates $x_{t,n}$ given by (\ref{eq:tx_coordinates}). Assuming analog beamforming, the weight of the $n$-th antenna is denoted as $w_n=e^{\mathrm{j}\phi_n}/\sqrt{N_t}$, where $\phi_n = \phi(x_{t,n}; B, F, \theta)$ denotes the phase shift sampled from the continuous phase profile in (\ref{BF}). The received electric field at $(z, x)$ is expressed as
\begin{equation}
    \label{discretize}
    E(z,x) = \frac{d}{\mathrm{j}\lambda} \sum_{n=1}^{N_t} w_n \frac{e^{\mathrm{j}k R_n}}{R_n} \cos\theta_n,
\end{equation}
where $R_n = \sqrt{(x - x_{t,n})^2 + z^2}$ represents the exact propagation distance and $\cos\theta_n = z/R_n$ is the directional cosine for the $n$-th antenna. Based on this discrete signal model, the received signal power is proportional to the squared magnitude of the electric field, i.e., $P_{rx} \propto |E(z_r, x_r)|^2$. 

To ensure the received performance at the Rx, the generated Airy beam trajectory must satisfy two geometric conditions: bypassing the obstacle and aligning with the Rx. To satisfy the first condition, we place a safe point $(z_b,x_{safe})$ beyond the upper or lower boundary of the obstacle, as depicted in Fig.~\ref{fig:system_model}. The transverse coordinate $x_{safe}$ is set to maintain a safe distance $d_{safe}$ from the selected boundary. This point guides the beam around the obstacle. Based on (\ref{trajectory}) and (\ref{trajectory peak}), to ensure that the bending beam trajectory passes through the safe point $(z_b, x_{safe})$, define $x_{01}$ as the aperture coordinate whose corresponding main-lobe peak is located at the safe point. Therefore, it must hold that
\begin{equation}
    \label{C1}
    \begin{cases}
        z_c(x_{01}) = z_b, \\
        x_{peak}(x_{01}) = x_{safe}.
    \end{cases} 
\end{equation}
 
Second, after bypassing the obstacle, the beam trajectory must pass through the Rx located at $(z_r, x_r)$. Similarly, we introduce $x_{02}$ as the aperture coordinate whose corresponding main-lobe peak is located at the Rx. Consequently, we have
\begin{equation}
    \label{C2}
    \begin{cases}
        z_c(x_{02}) = z_r, \\
        x_{peak}(x_{02}) = x_r.
    \end{cases} 
\end{equation}

To find the optimal parameter set, we aim to maximize the received signal power at the Rx subject to constraints (\ref{C1}) and (\ref{C2}). The resulting optimization problem is formulated as:
\begin{equation}
    \label{P1}
    \tag{P1}
    \begin{aligned}
    \boldsymbol{\Psi}_{\text{opt}} = \arg\max_{B, F, \theta} \quad & \left| \sum_{n=1}^{N_t}  \frac{e^{\mathrm{j}[\phi(x_{t,n}; B, F, \theta) + k R_n]}}{R_n} \cos\theta_n \right|^2 \\
    \text{s.t.} \quad & (10), \quad (11), \\
    & x_{01}, x_{02} \in [-L/2, L/2].
    \end{aligned}
\end{equation}

However, $\text{(P1)}$ appears to be highly non-convex and difficult to solve optimally. Although the authors in \cite{zhao2026airy} proposed a closed-form solution to $\text{(P1)}$, second-order focal-distance approximations are employed, which may degrade the communication performance under certain conditions, such as a large antenna aperture or a close user location, as will be shown via simulation in Section \ref{sec:results}.

\section{Proposed Optimization Framework}
\label{Optimization Framework}
In this section, we propose an alternative algorithm to solve $\text{(P1)}$ without relying on the second-order approximation. The proposed framework consists of geometric preprocessing and a two-layer optimization. First, geometric preprocessing determines the feasible range of $B$. Second, we develop a two-layer optimization framework. In the inner-layer problem, we adopt focal-curve intersections to recover the feasible values of $(F,\theta,x_{01},x_{02})$ for any given $B$ and return the largest received power to the outer problem, followed by the PS search on $B$.

\subsection{Geometric Preprocessing} 
First, we determine the feasible range of $B$. The launch angle must satisfy $|\sin\alpha(x_0)|\leq 1$. Since $\left|(F\sin\theta-x_0)/R_f(x_0)\right|\leq 1$ based on the definition of $R_f(x_0)$ in Section~\ref{sec:discretization}, (\ref{launch angle}) gives $\left|(2\pi B)^3x_0^2/k\right|\leq 2$. The maximum value of $x_0^2$ over the aperture occurs at $|x_0|=L/2$. Therefore, an upper bound on $|B|$ is 
\begin{equation}
    |B|\leq B_{\max} \triangleq \left(\frac{2}{\pi^2\lambda L^2}\right)^{1/3}.
    \label{Bmax}
\end{equation}
 
To avoid unnecessary computational cost of evaluating trajectories that bend toward the wrong side of the obstacle, we extend the line from the Tx origin through the obstacle's center, as shown in Fig.~\ref{fig:system_model}. The transverse coordinate of this line at $z=z_r$ is $x_{ref}=x_{obs}z_r/z_b$. By comparing $x_r$ with $x_{ref}$, the sign of $B$ is determined as
\begin{equation} 
\label{eq:sigma_B}
\sigma_B = \begin{cases}
+1, & \text{if } x_r \geq x_{ref}, \\
-1, & \text{otherwise},
\end{cases}
\end{equation}
and the safe point is selected as
\begin{equation}
x_{safe} = \begin{cases} 
x_{high} + d_{safe}, & \text{if } \sigma_B = +1,\\
x_{low} - d_{safe}, & \text{otherwise}.
\end{cases} 
\end{equation}
Thus, it must hold that $\sigma_BB>0$. By combining it with (\ref{Bmax}), the feasible range of $B$ can be determined.

\subsection{Proposed Solution to Inner-Layer Problem} 
\label{Proposed Solution to Inner-Layer Problem}
For a fixed $B$, the inner-layer problem determines ($F$, $\theta$, $x_{01}$, $x_{02}$) subject to constraints in (\ref{P1}). To this end, we first relax the constraint $x_{01}, x_{02} \in [-L/2, L/2]$, which allows $x_{01}$ and $x_{02}$ to lie outside the physical Tx aperture as virtual source coordinates. This relaxation is used to obtain candidate beam parameters, while the actual beamforming weights are still determined by the physical antenna positions in \eqref{eq:tx_coordinates}.

For notational simplicity, let $(z_1, x_1) = (z_b, x_{\text{safe}})$, $(z_2, x_2) = (z_r, x_r)$ and $\alpha_i \triangleq \alpha(x_{0i})$. Substituting the conditions $z_c(x_{0i}) = z_i$ and $x_{\text{peak}}(x_{0i}) = x_i, i=1,2$, specified by (\ref{C1}) and (\ref{C2}) into (\ref{emitted ray}) and (\ref{trajectory peak}) yields
\begin{equation}
    x_{0i}
    =
    x_i-z_i\tan\alpha_i
    -
    \frac{1.0188\lambda z_iB}{\cos^2\alpha_i}, i=1,2.
    \label{eq:inner_x0}
\end{equation}

Let $\beta=(2\pi B)^3/k$ and $u_i=\sin\alpha_i-\beta x_{0i}^2$. Combining the launch-angle relation in (\ref{launch angle}) with the caustic condition $z_c(x_{0i})=z_i$, we can express the coordinate of the focal point as:  
\begin{equation}
    \begin{cases}
    F\cos\theta
    =\dfrac{(1-u_i^2)^{3/2}}
    {2\beta x_{0i}+\cos^3\alpha_i/z_i}\triangleq  p_i(\alpha_i),\\[3pt]
    F\sin\theta
    =x_{0i}+\dfrac{p_i(\alpha_i)u_i}{\sqrt{1-u_i^2}}\triangleq  q_i(\alpha_i).
    \end{cases}
    \label{eq:focus_curve}
\end{equation}

As $\alpha_i$ changes within its feasible range, $(p_1(\alpha_1),q_1(\alpha_1))$ and $(p_2(\alpha_2),q_2(\alpha_2))$ form two distinct curves, and their intersection gives a common focal point $(F\cos\theta,F\sin\theta)$ that satisfies both (\ref{C1}) and (\ref{C2}). We therefore solve
\begin{equation}
    \bm{r}(\alpha_1,\alpha_2)
    =
    \begin{bmatrix}
        p_1(\alpha_1)
        -
        p_2(\alpha_2)
        \\[2mm]
        q_1(\alpha_1)
        -
        q_2(\alpha_2)
    \end{bmatrix}
    =
    \bm{0}
    \label{eq:curve_intersection}
\end{equation}  
by sampling the two curves and locating their intersections. Let $(p,q)$ denote an intersection point. Then, the focal distance and steering angle can be obtained as $F=\sqrt{p^2+q^2}$ and $\theta=\operatorname{atan2}(q,p)$. Meanwhile, substituting $\alpha_1$ and $\alpha_2$ into (\ref{eq:inner_x0}) yields $x_{01}$ and $x_{02}$ for the target positions $(z_b, x_{\text{safe}})$ and $(z_r, x_r)$, respectively. Note that if either $x_{01}$ or $x_{02}$ lies outside $[-L/2,L/2]$, it is retained only as a virtual coordinate introduced by the relaxation. The corresponding parameter values $(B,F,\theta)$ are still used to construct the phase profile for the antenna array. If multiple valid intersections are found, we evaluate the received power for each solution and retain the one with the highest value for the outer-layer search.

\subsection{Outer PS Search over $B$}  
Let \(g(B)\) denote the optimal objective value of (\ref{P1}) for a given \(B\), obtained via the inner-layer solution. Due to the non-convexity of $g(B)$, we conduct a 1D search over the feasible range of \(B\) using the PS strategy \cite{gokcesu2024cumulative}. We first uniformly sample the search range and estimate the Lipschitz constant $\widehat{L}_g$ from the maximum absolute slope between adjacent sampled points. The PS method then uses this estimate to select new sampling points and progressively refine the search. The detailed iterative steps can be found in \cite{gokcesu2024cumulative}. Finally, the sampled point with the highest \(g(B)\) is selected as \(B_{\mathrm{opt}}\). The corresponding \(F_{\mathrm{opt}}\) and \(\theta_{\mathrm{opt}}\) are obtained from the inner-layer solution.

\section{Simulation Results}
\label{sec:results}
\begin{figure*}[!b] 
    \centering
    \begin{subfigure}[b]{0.49\textwidth} 
        \centering
        \includegraphics[width=\textwidth]{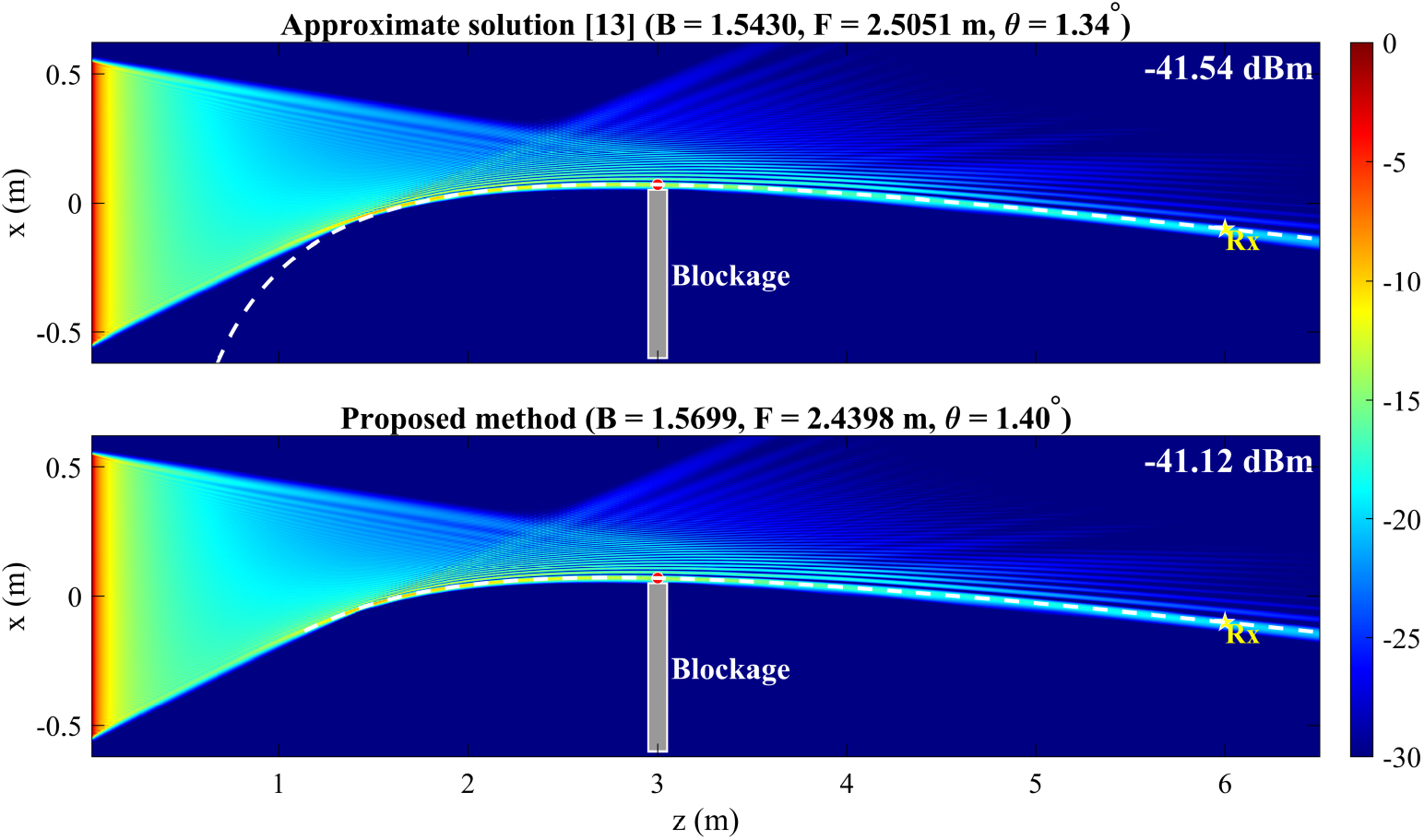}
        \caption{Normal near-field distance.}
        \label{fig:sim_paraxial}
    \end{subfigure} 
    \hfil 
    \begin{subfigure}[b]{0.49\textwidth}
        \centering
        \includegraphics[width=\textwidth]{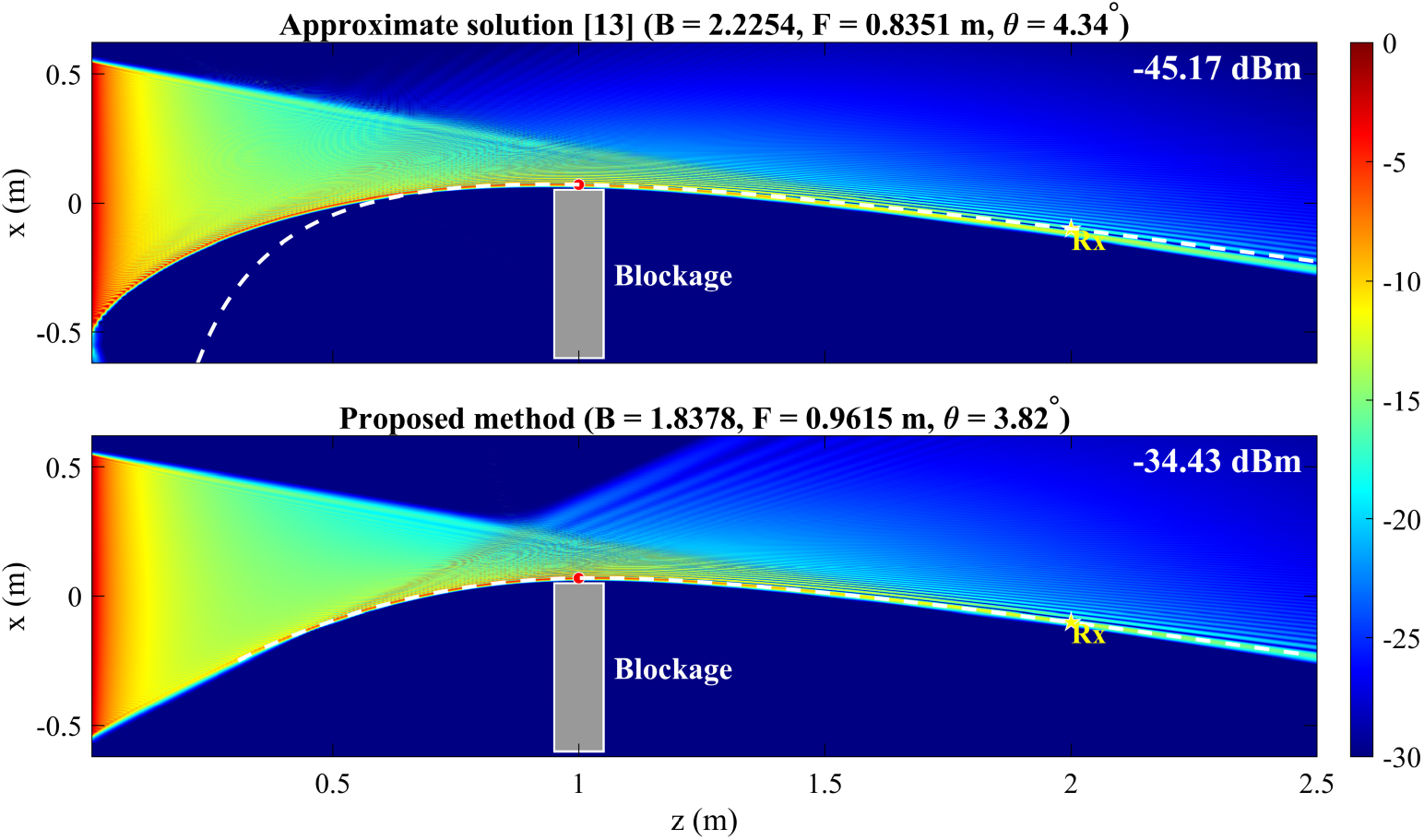}
        \caption{Short near-field distance.}
        \label{fig:sim_large_angle}
    \end{subfigure}
    
    \caption{Comparison of spatial field distributions for different schemes.} 
    \label{fig:all_simulations}  
\end{figure*}
In this section, simulation results are provided to demonstrate the efficacy of the proposed exact method compared to the following two baseline schemes. The first baseline is the closed-form solution presented in \cite{zhao2026airy}, which is based on a second-order Taylor approximation of the focal distance. The associated phase profile is given by
\begin{equation}
    \phi_n^{\text{appr}} = \frac{(2\pi B)^3}{3}x_{t,n}^3 - \frac{k}{2F}x_{t,n}^2 - k\sin\theta x_{t,n}.
\end{equation}

The second baseline is the conventional near-field focused beam (FB) at the Rx, and its phase shifts are given by 
\begin{equation}
    \phi_n^{\text{FB}} = -k\sqrt{(x_r - x_{t,n})^2 + z_r^2}.
\end{equation}

Unless otherwise stated, the simulation parameters are set as follows. The carrier frequency is 140 GHz. The Tx is equipped with $N_t=1024$ elements with spacing $d=\lambda/2$. The near-field channel with blockage is generated using the cascaded geometrical and wave channel model in \cite{zhao2025terahertz}. The transmit power is $P_{\rm tx}=30$ dBm. We define the LoS blockage ratio as $\rho_{\mathrm{LoS}}=N_{\mathrm{blocked}}/N_t$, where $N_{\mathrm{blocked}}$ is the number of direct paths from the Tx antennas to the Rx that intersect the rectangular obstacle. A path is considered blocked if it intersects any part of the obstacle.

In the simulation, we first compare the spatial field distributions of the approximate solution in \cite{zhao2026airy} and the proposed method in two representative scenarios with normal and short propagation distances, respectively, as shown in Fig.~\ref{fig:all_simulations}. In both scenarios, we set the transverse obstacle boundaries defined in \eqref{eq:obstacle_mask} as $x_{\rm low}=-0.60$ m and $x_{\rm high}=0.05$ m, with a safety margin of $d_{\rm safe}=0.02$ m. In Fig.~\ref{fig:all_simulations}(\subref{fig:sim_paraxial}), the obstacle extends from $z=2.95$ m to $z=3.05$ m, and the Rx is located at $(z_r,x_r)=(6.0,-0.1)$ m, with $\rho_{\mathrm{LoS}}\approx68.65\%$. In Fig.~\ref{fig:all_simulations}(\subref{fig:sim_large_angle}), the obstacle extends from $z=0.95$ m to $z=1.05$ m, and the Rx is located at $(z_r,x_r)=(2.0,-0.1)$ m, with $\rho_{\mathrm{LoS}}\approx69.63\%$.

It is observed from Fig.~\ref{fig:all_simulations}(\subref{fig:sim_paraxial}) that the Fresnel approximation remains reasonably accurate. Both methods form a curved beam that passes around the obstacle and reaches the Rx, with received powers of $-41.54$ dBm for the approximate solution and $-41.12$ dBm for the proposed method. However, as observed from Fig.~\ref{fig:all_simulations}(\subref{fig:sim_large_angle}), the Fresnel approximation breaks down at such short distances. The approximate solution shows clear beam defocusing and its received power decreases to $-45.17$ dBm. The proposed method follows the intended caustic more closely and achieves a received signal power of $-34.43$ dBm at the Rx, which is $10.74$ dB higher than the approximate solution. These results show that the Fresnel approximation is sufficient in the normal near-field regime but becomes inaccurate as the propagation distance becomes smaller.

\begin{figure}[!t]
    \centering
     \includegraphics[width=0.82\columnwidth]{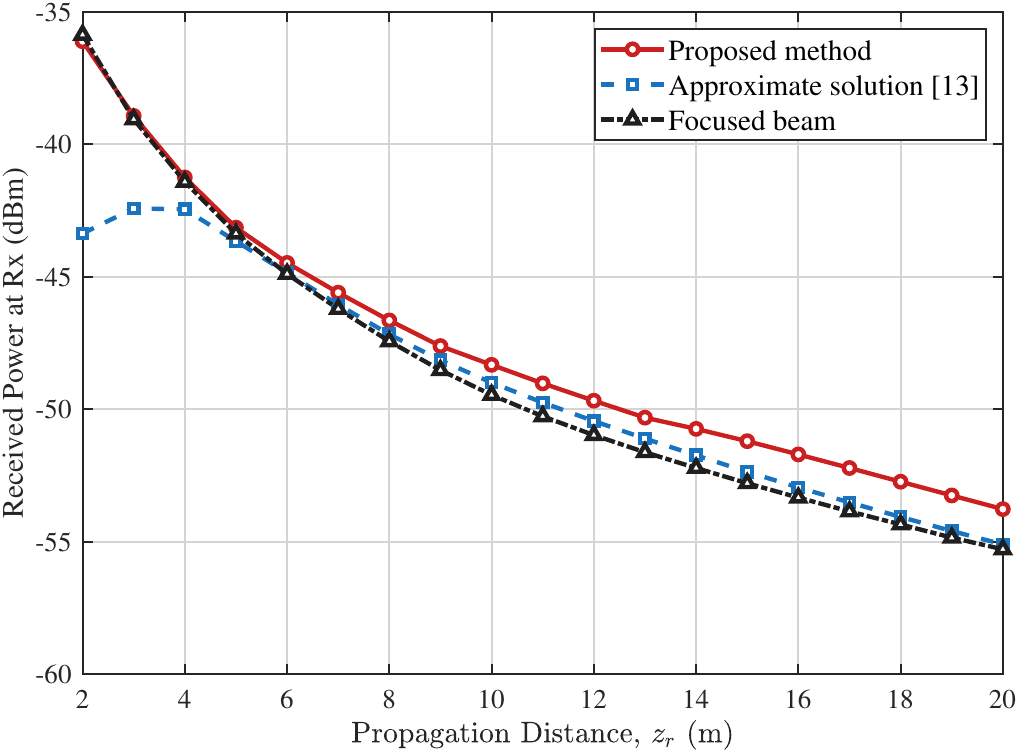}
    \caption{Received power versus propagation distance under an $80\%$ LoS blockage ratio.}
    \label{propagation_distance}
\end{figure}

Next, Fig.~\ref{propagation_distance} compares the performance of different schemes as $z_r$ varies from $2$ m to $20$ m with $x_r=-0.1$ m. The obstacle is centered at $z_r/2$ with a fixed thickness of $0.1$ m. The lower boundary of the obstacle remains fixed, while the upper boundary is adjusted to maintain $\rho_{\mathrm{LoS}}=80\%$. As $z_r$ increases, the received powers generally decrease due to more significant propagation loss. The performance gap between the proposed and approximate solutions first narrows, reaching a minimum at around $z_r=6$ m, before widening again. The initial narrowing is mainly due to reduced Fresnel approximation error. At longer distances, however, parameter selection becomes the dominant factor, as different beam parameters produce different field distributions around the obstacle, resulting in greater power loss for the approximate solution. It is also observed that the proposed method achieves an increasing received-power advantage over FB as $z_r$ increases.  

\begin{figure}[!t]
    \centering
    \includegraphics[width=0.82\columnwidth]{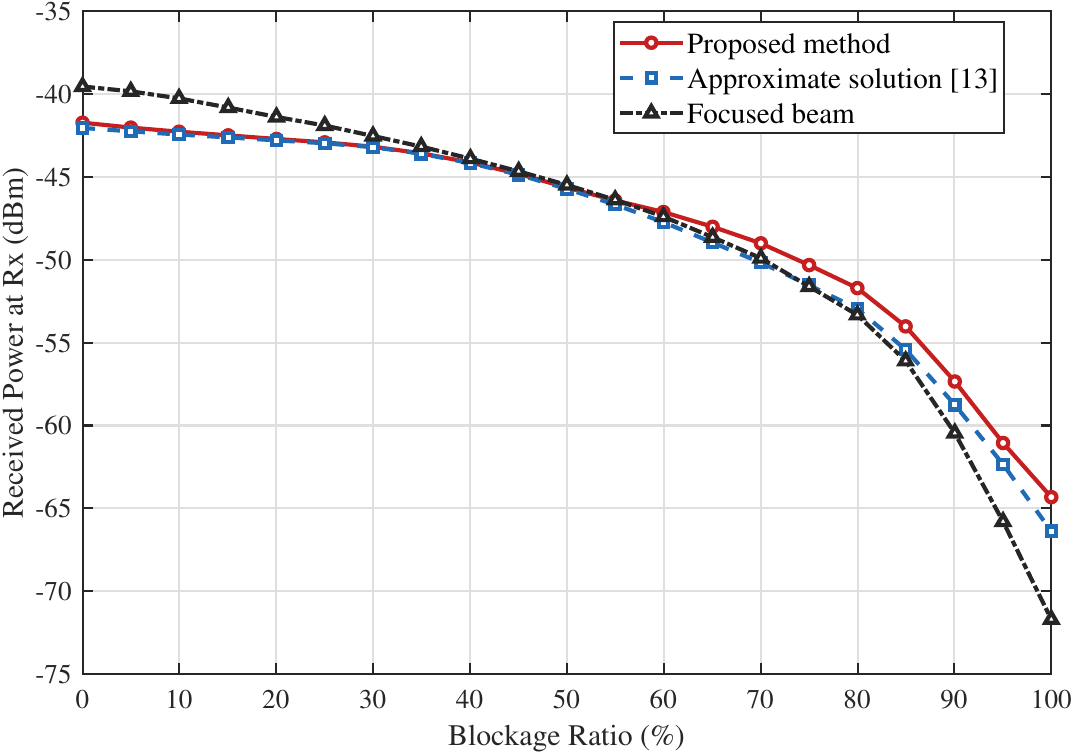} 
    \caption{Received power versus blockage ratio with the Rx located at $z_r=16.0$ m.}
    \label{traditional}
\end{figure}

Fig.~\ref{traditional} compares the received powers of the three methods under different blockage ratios. The Rx is located at $(z_r,x_r)=(16.0,-0.1)$ m and the obstacle is centered at $z_r/2$ with a fixed thickness of $0.1$ m. The blockage ratio increases by extending the upper obstacle boundary while keeping the lower boundary fixed. For small LoS blockage ratios, FB is observed to achieve the highest received power as it concentrates the transmitted field at the Rx. The two Airy-beam-based methods perform slightly worse because their energy is distributed along curved trajectories. As the blockage ratio increases, FB power initially decreases slowly as some direct paths remain unblocked, but drops sharply when most are blocked. In contrast, the proposed method guides the beam around the obstacle and surpasses FB beyond a blockage ratio of $60\%$. The approximate solution shows a similar trend but generally achieves slightly lower received power than the proposed method.

\begin{figure}[!t]
    \centering
    \includegraphics[width=0.82\columnwidth]{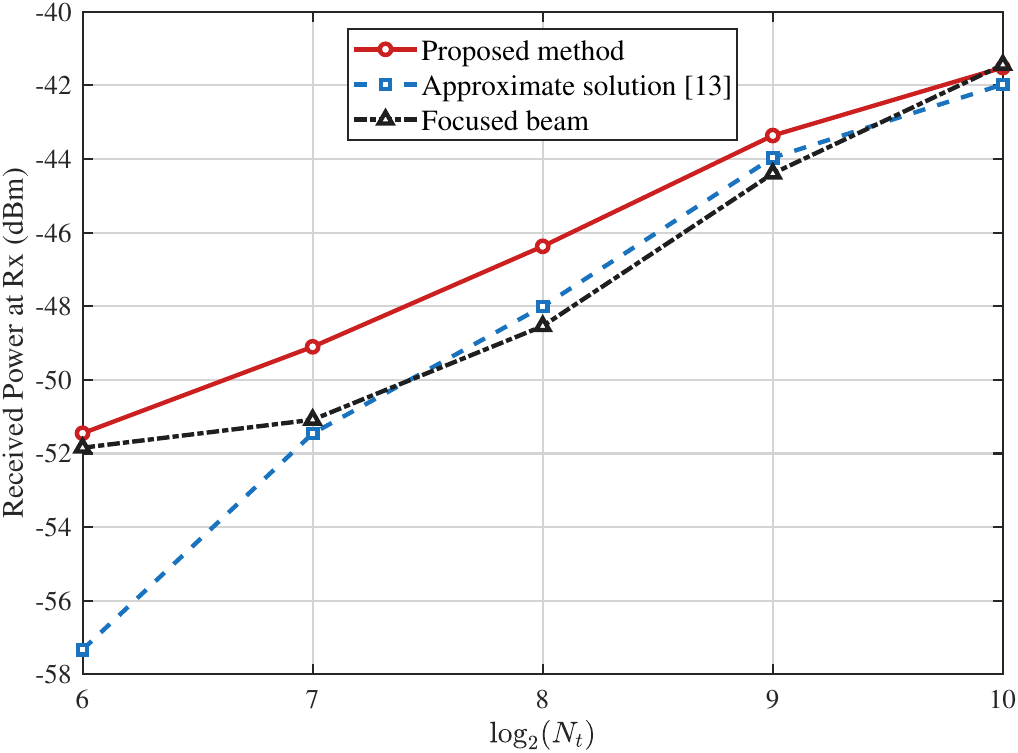} 
    \caption{Received power versus Tx array size under a $70\%$ LoS blockage ratio.} 
    \label{antenna_count} 
\end{figure}

Finally, Fig.~\ref{antenna_count} shows the received signal power versus the logarithm of the number of Tx antennas, $\log_2(N_t)$, under a fixed LoS blockage ratio of $70\%$, by adjusting the upper obstacle boundary for each $N_t$. The Rx is located at $(z_r,x_r)=(5.0,-0.1)$ m, and the obstacle extends from $z=3.45$ m to $z=3.55$ m. It is observed that as $N_t$ increases, the received signal powers of all considered schemes improve. The proposed method outperforms both baselines for smaller arrays, but FB gradually closes the gap as $N_t$ increases. This is because more unblocked antennas become available under a fixed blockage ratio, allowing FB to concentrate more energy directly at the Rx. In contrast, the proposed method distributes energy along a curved trajectory, leading to lower energy focused to the Rx.   

\section{Conclusion} 
In this letter, we proposed an exact Airy-beamforming optimization method for blockage avoidance in near-field THz communications. By retaining the exact focal-distance expression and using the corresponding caustic geometry, the method determines the beam parameters through focal-curve intersections and PS search. Simulation results showed that the proposed method preserves the intended beam trajectory when the Fresnel approximation becomes inaccurate, achieving higher received signal power than the existing approximate design. Furthermore, it outperforms conventional focused beams under severe blockage by effectively steering energy around obstacles.

\bibliographystyle{IEEEtran}
\bibliography{reference}

\end{document}